\documentclass[sigconf,authorversion,nonacm]{acmart}

\AtBeginDocument{%
  \providecommand\BibTeX{{%
    \normalfont B\kern-0.5em{\scshape i\kern-0.25em b}\kern-0.8em\TeX}}}

\setcopyright{acmlicensed}
\copyrightyear{2025}
\acmYear{2025}
\acmDOI{XXXXXXX.XXXXXXX}

\usepackage{makecell}
\usepackage{multirow}
\usepackage{booktabs}
\usepackage{array}
\usepackage[normalem]{ulem}

\usepackage{multicol}
\usepackage{cuted}
\usepackage[font=small,labelfont=bf,tableposition=top]{caption}
\usepackage{enumitem}

\usepackage{cellspace}
\usepackage{tabularx}

\begin{document}

\title[Motivation and Design Opportunities in Group Therapy]{When Group Spirit Meets Personal Journeys: \\Exploring Motivational Dynamics and Design Opportunities\\ in Group Therapy
} 


\author{Shixian Geng}
\authornote{Both authors contributed equally to this research and shared the first authorship.}
\affiliation{%
  \institution{The University of Tokyo}
  \streetaddress{7-3-1 Hongo}
  \city{Bunkyo-ku}
  \state{Tokyo}
  \country{Japan}
  \postcode{113-8656}
}
\email{shixiangeng@iis-lab.org}

\author{Ginshi Shimojima}
\authornotemark
\affiliation{%
  \institution{The University of Tokyo}
  \streetaddress{7-3-1 Hongo}
  \city{Bunkyo-ku}
  \state{Tokyo}
  \country{Japan}
  \postcode{113-8656}
}
\email{ginshi@iis-lab.org}

\author{Chi-Lan Yang}
\affiliation{%
  \institution{The University of Tokyo}
  \streetaddress{7-3-1 Hongo}
  \city{Bunkyo-ku}
  \state{Tokyo}
  \country{Japan}
  \postcode{113-8656}
}
\email{chilan.yang@iii.u-tokyo.ac.jp}

\author{Zefan Sramek}
\affiliation{%
  \institution{The University of Tokyo}
  \streetaddress{7-3-1 Hongo}
  \city{Bunkyo-ku}
  \state{Tokyo}
  \country{Japan}
  \postcode{113-8656}
}
\email{zefans@iis-lab.org}

\author{Shunpei Norihama}
\affiliation{%
  \institution{The University of Tokyo}
  \streetaddress{7-3-1 Hongo}
  \city{Bunkyo-ku}
  \state{Tokyo}
  \country{Japan}
  \postcode{113-8656}
}
\email{norihama@iis-lab.org}

\author{Ayumi Takano}
\affiliation{%
  \institution{National Center of Neurology and Psychiatry}
  \streetaddress{4 Chome-1-1 Ogawahigashicho}
  \city{Kodaira-shi}
  \state{Tokyo}
  \country{Japan}
  \postcode{187-8551}
}
\email{atakano@ncnp.go.jp}

\author{Simo Hosio}
\affiliation{%
  \institution{University of Oulu}
  \streetaddress{Pentti Kaiteran katu 1}
  \city{Oulu}
  \country{Finland}
  \postcode{90570}
}
\email{simo.hosio@oulu.fi}

\author{Koji Yatani}
\affiliation{%
  \institution{The University of Tokyo}
  \streetaddress{7-3-1 Hongo}
  \city{Bunkyo-ku}
  \state{Tokyo}
  \country{Japan}
  \postcode{113-8656}
}
\email{koji@iis-lab.org}


\begin{abstract}
Psychotherapy, such as cognitive-behavioral therapy (CBT), is effective in treating various mental disorders. Technology-facilitated mental health therapy improves client engagement through methods like digitization or gamification. However, these innovations largely cater to individual therapy, ignoring the potential of group therapy—a treatment for multiple clients concurrently, which enables individual clients to receive various perspectives in the treatment process and also addresses the scarcity of healthcare practitioners to reduce costs. Notwithstanding its cost-effectiveness and unique social dynamics that foster peer learning and community support, group therapy, such as group CBT, faces the issue of attrition. While existing medical work has developed guidelines for therapists, such as establishing leadership and empathy to facilitate group therapy, understanding about the interactions between each stakeholder is still missing. To bridge this gap, this study examined a group CBT program called the Serigaya Methamphetamine Relapse Prevention Program (SMARPP) as a case study to understand stakeholder coordination and communication, along with factors promoting and hindering continuous engagement in group therapy. In-depth interviews with eight facilitators and six former clients from SMARPP revealed the motivators and demotivators for \textit{facilitator-facilitator}, \textit{client-client}, and \textit{facilitator-client communications}. Our investigation uncovers the presence of discernible conflicts between clients' intrapersonal motivation as well as interpersonal motivation in the context of group therapy through the lens of self-determination theory. We discuss insights and research opportunities for the HCI community to mediate such tension and enhance stakeholder communication in future technology-assisted group therapy settings.
\end{abstract}





\maketitle

\section{Introduction}

Modern psychotherapy, such as cognitive-behavioral therapy (CBT), has been effective for various mental health issues such as anxiety disorders~\cite{heimberg2002cognitive}, mood disorders~\cite{fava1998prevention}, substance use disorders~\cite{kaminer1998psychotherapies}, and eating disorders~\cite{fairburn2008cognitive}. 
\textit{Group therapy} like group CBT provides opportunities for more clients to receive proper treatment at a much lower cost and with a need for fewer therapists~\cite{himle2003group}. It involves multiple stakeholders, including facilitators, recovered members, senior members, and newcomers. Related work has revealed that the interactions and dynamics within a well-organized CBT group can positively influence its members in many ways~\cite{bieling2022cognitive}. For instance, a member can gain community support and learn from others as they observe and discuss with others who share the same issues and mindset. 
From the therapist's perspective, observing the interpersonal interactions between members serves as useful information to reveal a client's cognitive and behavioral patterns~\cite{bieling2022cognitive}.
Although group CBT shows great potential for clients' recovery, the dropout rate of such therapy is still high~\cite{fernandez2015meta}.

Digital psychotherapy, the use of digital tools and devices to deliver or improve mental and behavioral therapy, is now a thriving research area with over three decades of clinical and academic work behind it~\cite{thew2022advances}. 
For example, digital technologies can enable remote CBT treatment on smartphones or web apps~\cite{andersson2014internet}. Some works integrate gamification to encourage treatment~\cite{coyle2011exploratory,tochilnikova2022guilty,assigana2014tf}. AIs and virtual agents are applied to understand user behaviors and encourage behavior change~\cite{calvaresi2019social,goonesekera2022cognitive,beilharz2021development}. Virtual Reality (VR) can also help synthesize the preferred environment for treatment and conduct exposure therapy~\cite{wallach2009virtual,lindner2021better}.
However, the existing technology design has primarily focused on 1-on-1 psychotherapeutic settings, without substantially investigating its group form.
While the practical clinical implementation of group therapy, such as group CBT, is rather well understood~\cite{bieling2022cognitive}, there exists a lack of understanding of the coordination and communication process between different stakeholders in the modern media environment, as well as understanding of the current usage of technology in such group therapy. In addition, while continued attendance is key for client recovery~\cite{persons1988predictors}, it remains unclear how technology can be applied to motivate many clients collectively in group therapy for consistent engagement and continued attendance.
We argue a better understanding of these issues is pivotal for designing collaborative health technologies for group therapy.

To establish an understanding of motivational dynamics in group therapy to explore potential technological interventions that enhance clients' motivation, we investigate the communication practices between different stakeholders in group therapy, as relational factors are distinctive features of group therapy compared to individual therapy. We set out to answer the following research questions through a qualitative study of the Serigaya Methamphetamine Relapse Prevention Program (SMARPP), an existing, well-established group therapy program for substance use disorders in Japan~\cite{matsumoto2015treatment}: 

\begin{enumerate}[label=\textbf{RQ.\arabic*}]
    \item \textbf{What are the motivators and demotivators a client faces when communicating with other stakeholders in group therapy?}
    \item\textbf{When and how does each stakeholder in group therapy coordinate and communicate with each other, and how does the communication relate to the client's motivation?}

\end{enumerate}

We conducted semi-structured interviews with 8 facilitators and 6 former clients who have participated in SMARPP. 
Our results uncover the varied landscape of stakeholder communication, highlighting the importance and challenge of facilitators providing personalized empathy and clients maintaining appropriate social boundaries, and providing insight into the value associated with both in-person and online group therapy. To analyze our results further through theoretical motivational frameworks, we present an in-depth exploration of HCI research opportunities by drawing insights from Self-Determination Theory (SDT). By incorporating SDT into our findings, we aim to reveal the motivational dynamics within the group therapy context, understanding how technology can be applied to enhance the collective motivation of clients in the group.
Given the limited research in HCI and CSCW focused on group therapy, our work offers an outline of its key features and challenges, and proposes avenues for future exploration of collaborative technology solutions to support group therapy, both from the perspective of clients and facilitators, which encourages further research on group therapy within the HCI and CSCW community. 

The contributions of this work are summarized as follows:

\begin{enumerate}
    \item An analysis of semi-structured interviews with the direct stakeholders, including facilitators and clients, of a group CBT program for substance use disorder
    
    \item An empirical understanding of motivators and demotivators influencing continued attendance of clients
    and of communication that occurs before, during, and after the program sessions between different stakeholders
    
    \item A discussion of the intrapersonal and interpersonal motivational tensions existing in the group therapy context through the lens of Self-Determination Theory 

    \item Research opportunities for future collaborative technology in HCI and CSCW to mediate the motivational tensions existing in the group therapy context

\end{enumerate}

\section{Related Work}
\subsection{Cognitive-Behavioral Therapy in HCI}
\label{sec: cbt in HCI}
Cognitive-behavioral therapy (CBT)~\cite{beck2020cognitive}stands as one of the most commonly used psychotherapies for effectively treating various mental health disorders. It involves the process of recognizing external triggers and internal maladaptive thoughts that contribute to negative emotions and irrational behavioral responses, followed by a methodical and unbiased analysis of these situations~\cite{beck2020cognitive}.
As a result, CBT typically comprises two components: the \textit{didactic process} aimed to educate clients on psychological knowledge and techniques, and \textit{experiential learning} involving therapeutic discussions and exercises designed to help clients assess their thoughts and behaviors~\cite{bieling2022cognitive}.

Researchers have carried out empirical investigations to gain insights into the effectiveness of online 1-on-1 CBT programs~\cite{ruwaard2012effectiveness,spence2011randomized,andersson2008pros,ruwaard2011efficacy}, organized workshops to develop digital applications that can enhance such CBT sessions~\cite{thieme2023designing,tochilnikova2022guilty}, and explored the perceived empathy of a support seeker toward a helper in online CBT-based peer support platforms~\cite{syed2024machine}.
Accordingly, HCI researchers have conducted extensive research on technology-supported CBT, covering a wide spectrum of disorders such as anxiety~\cite{coyle2011exploratory,thieme2023designing}, depression~\cite{thieme2023designing}, panic disorder~\cite{lee2018moonglow}, autism~\cite{giusti2011dimensions}, and addiction~\cite{carroll2014computer}.

Despite the comprehensive coverage of topics and interactions, previous HCI work has mainly focused on 1-on-1 CBT, and has only minimally addressed CBT delivered in group settings.
For example, Lederman et al. integrated social networking into web-based mental health therapy in the creation of a social therapy environment~\cite{lederman2014moderated}, while Gorini et al. evaluated the use of 3D virtual worlds to create virtual communities of patients~\cite{gorini2008second}.
However, these studies mainly focused on proposing and validating specific designs that empower the patient community instead of supporting group therapy, and the process and challenges of group therapy are left unexamined. 
Although there are well-established theoretical guidelines from the therapists' perspectives for offering group CBT~\cite{bieling2022cognitive}, there is a lack of investigation of the communication dynamics among group members and their influence on the therapeutic process, particularly from the HCI standpoint. Specifically, the integration of technology into group therapy remains an unexplored avenue within the HCI community. There is a lack of digital tools to support group therapy and a lack of clear design principles for the development of such tools.

Our study addresses this gap by exploring the communications and unique challenges associated with group CBT and by discussing research opportunities for effectively supporting such group therapy with technology. Based on our understanding of the motivational dynamics in group therapy, we explore possibilities for digital tools that support facilitators' effective management of group dynamics, and tools that mitigate conflicts and promote client motivation.

\subsection{Motivation Studies in HCI with Self-Determination Theory}
Sustained motivation is the key factor in fostering commitment and engagement to prevent clinical attrition~\cite{bados2007efficacy}. Studying motivation in HCI is crucial for the development of digital interventions that enhance user engagement in a particular system. While current technological solutions for mental health address individual motivation, they may not fully account for the intricate social dynamics presented in group settings that may influence collective motivation. To bridge this gap, our work explores the application of Self-Determination Theory (SDT)~\cite{deci2013intrinsic} as a framework for understanding participants' motivation in group therapy. Poeller et al. highlighted the importance of evaluating and selecting appropriate motivation theories in HCI studies~\cite{poeller2022self}, and we decided to choose SDT over other motivational theories due to its focus on promoting intrinsic motivation, motivation coming from intrinsic satisfaction when engaging in a task rather than external rewards or pressures~\cite{deci2013intrinsic}. This aligns with our research context of supporting mental well-being. Additionally, SDT's emphasis on autonomy and relatedness is particularly relevant in the context of group therapy, where interpersonal relationships and individual autonomy play crucial roles.

According to SDT, individuals have three basic psychological needs that drive their motivation: \textbf{Competence}, the capacity to meet and overcome challenges effectively; \textbf{Autonomy}, the liberty to make personal choices; and \textbf{Relatedness}, the social connection with others. When individuals find themselves within an environment conducive to fulfilling these three psychological needs, there is a propensity for higher motivation, fostering persistent engagement in their activities~\cite{ryan2000self}.

Within the HCI community, SDT is also an increasingly popular framework frequently used in investigations on user motivation in gaming, education, and health research. For example, SDT has served as a theoretical foundation guiding the design of various technologies, including conventional agents~\cite{yang2021designing}, social robots~\cite{van2020using}, fitness applications~\cite{villalobos2021informed}, and virtual persona creation~\cite{jansen2017personas}. In these instances, researchers aimed to design technology that supports the three basic psychological needs of its users. SDT has also been employed as a theoretical framework to analyze human behaviors in HCI and CSCW, such as the motivation of young women studying Computer Science~\cite{mishkin2019applying} and the basic needs of older adults using technology during the COVID-19 pandemic~\cite{zhao2023older}. In the context of digital health most relevant to this study, Aufheimer et al. conducted an interview study with participants in physical therapy and offered SDT-based recommendations on game design for physical therapy~\cite{aufheimer2023examination}.

However, previous HCI studies primarily concentrate on using SDT to analyze the basic psychological needs for \textit{individual} users, irrespective of whether the user is part of a group or a community. Such focus overlooks the intricate social and motivational dynamics that occur among diverse stakeholders and their potential impact on each other's psychological needs and motivation. 
Recent work in HCI has underscored the necessity for a more comprehensive and diverse exploration of SDT and its associated sub-theories~\cite{tyack2020masterclass,ballou2022self,tyack2020self}. While previous studies have explored individual differences in motivation using the SDT sub-theory Causality Orientations Theory~\cite{mekler2017towards,wang2012fine}, or demonstrated the dynamic and fluctuating features of motivation and needs~\cite{pesonen2022weekly}, these investigations fall short of addressing the motivational challenges and conflicts of all stakeholders in collaborative tasks. 
Our research endeavors to fill this gap by examining the interpersonal influences on self-determination within a typical group context in healthcare: group therapy.

\subsection{Social Factors in Group Therapy and SMARPP}

Group therapy demonstrates similar effectiveness and unique benefits compared to individual therapy, such as fostering relatability, reducing stigma, and facilitating vicarious learning~\cite{bieling2022cognitive}.
Yalom
identified a series of therapeutic factors that contribute to the efficacy of group therapy, including the instillation of hope from recovered members, the sense of not being alone, and group cohesiveness~\cite{yalom2020theory}.
Consequently, relational factors, such as communication and interactions between group members together with the therapists, are distinctive components of group therapy compared to individual therapy.

However, prior literature has also identified challenges associated with group therapy.
Therapeutic discussions can be hindered by dominating members or negative influencers~\cite{yalom2020theory}, and the diverse therapeutic needs and cognitive levels of group members pose challenges to the therapist's group management skills~\cite{bieling2022cognitive}.
Medical literature tends to address these challenges primarily through interventions by therapists; Guidelines for therapists are designed to improve their characteristics, such as leadership and empathy in managing specific issues in groups~\cite{yalom2020theory,bieling2022cognitive}.
While the HCI community has extensively explored technological interventions for individual therapy as mentioned in Section~\ref{sec: cbt in HCI}, technological interventions for group therapy are still under-explored.
Group interactions and communications are unique factors in group therapy.
Thus, to bridge this gap, our study aims to explore the motivation dynamics of clients by examining stakeholder communications in group therapy and uncovering the causality of existing challenges within the social context of group therapy.

We chose the Serigaya Methamphetamine Relapse Prevention Program (SMARPP)~\cite{matsumoto2015treatment,kobayashi2007preliminary} as a case study, because it is a well-established evidence-based group therapy program in Japan and is covered by national health insurance for substance use disorder.
SMARPP was developed based on the Matrix Model for outpatient stimulant abuse treatment~\cite{rawson1995intensive,kobayashi2007preliminary}. The Matrix Model is a structured framework developed in the 1980s at the Matrix Institute on Addiction in Los Angeles, California~\cite{shoptaw1995matrix}.
It involves a mix of evidence-based approaches of CBT and motivational interviewing~\cite{hettema2005motivational}. 
SMARPP facilitators are trained through workshops to manage groups following group therapy practices~\cite{yalom2020theory}.
Although SMARPP primarily focuses on substance use, including methamphetamine, marijuana, prescription drugs, and alcohol, its clients often present comorbid conditions of other mental issues.
These individuals are frequently marginalized by society and subjected to heavily stigmatized and biased attitudes from others, which can potentially trigger their substance use.
Given SMARPP's incorporation of various evidence-based approaches from both Japanese and Western literature, we consider SMARPP to be a representative program to inform general group therapy.

Previous studies have investigated web-based SMARPP for e-learning, or e-SMARPP~\cite{takano2016web,takano2020effect}. These studies focused on the digitalization of the psychoeducation portions of therapy for individual use, rather than offering it in a group format as is done in traditional in-person sessions.
However, it was found that e-SMARPP failed to demonstrate efficacy compared to the self-monitoring control group~\cite{takano2020effect}.
Thus, it is worth exploring how future technology could adapt to the unique dynamics of group therapy environments, going beyond mere digitalization of content to leverage the communicational advantages unique to group therapy and enhance treatment efficacy.

\section{Qualitative Examinations on Communication Practices and Challenges in SMARPP}

To develop an understanding of communication practices and the experience of group psychotherapy, we conducted qualitative investigations with two main stakeholders of SMARPP: facilitators and clients. 

\begin{table*}[t]
\small
    \begin{minipage}[t]{0.5\textwidth}
        \centering
        \caption{Demographic data of facilitator participants.}
        \begin{tabular}{cccc}
           \toprule[1pt]\midrule[0.3pt]
           ID  &   Gender &  Age & Years in SMARPP\\
           \midrule
            PF1 & Female & 40s & 18 \\
            PF2 &  PNS & 30s &  9\\
            PF3 &  Female &  40s&  12\\
            PF4 &  Female &  40s &  12\\
            PF5 &  PNS &  40s &  10\\
            PF6 &  Male &  30s &  6\\
            PF7 &  Female & 50s & 15\\
            PF8 &  Male & 40s &  10\\
           \bottomrule[1pt]
        \end{tabular}
        \label{tab:d_f}
    \end{minipage}
    \begin{minipage}[t]{0.48\textwidth}
        \centering
        \caption{ Demographic data of client participants.}
        \begin{tabular}{cccc}
           \toprule[1pt]\midrule[0.3pt]
            ID  &  Gender &  Age &  Years in SMARPP\\
           \midrule
            PC1 & Female &  40s &  5 \\
            PC2 &  Female &  50s &  10 \\
            PC3 & Male &  60s &  1.5 \\
            PC4 &  Male &  40s &  8 \\
           PC5 &  Female & 50s &  1 \\
            PC6 &  Male &  30s &  9 \\
          \bottomrule[1pt]
        \end{tabular}
        \label{tab:d_p}
    \end{minipage}
\end{table*}

\subsection{Structure of SMARPP}
\label{sec:smarpp structure}
In SMARPP, clients meet in person once a week for 90 minutes with a facilitator, a co-leader to assist the facilitator, and a peer staff member to represent people who have recovered from addiction via the program.
10--15 clients typically join a gathering. 
The facilitator leads the program.
The entire gathering is divided into three sessions: check-in, psychoeducation with workbook reading, and check-out.
The check-in session, typically the longest session, is for discussing each client's activities and substance use in the past week.
Each participant typically has 3--5 minutes to speak, though other clients are welcome to make comments.
In the workbook reading session, clients learn about symptoms of addiction, triggers of cravings, and coping strategies, and discuss with other clients and staff.
This session typically lasts 30 minutes, depending on how long the check-in session was.
The checkout session focuses on each client making preparations for the upcoming weeks.
For example, they review situations where cravings are likely to occur and plan coping strategies.
This session can be short or even omitted depending on the flow of two previous sessions.

SMARPP facilitators rotate shifts for each meeting on different weekdays due to limited numbers of staff.
To promote participation and flexibility, clients are welcome to join any meeting, and it is acceptable for newcomers not to start from the beginning of the program.
Clients also have the freedom to miss or drop out of sessions.
This organization policy often results in varying group compositions at each gathering, which can potentially influence the group dynamic of each session.

\subsection{Study Participants}
We recruited eight SMARPP facilitators with experience conducting in-person SMARPP programs (referred to as facilitator participants, PF1 -- PF8).
All facilitator participants have served as SMARPP facilitators at the same particular medical institution in Japan.
The interviews with them were intended to uncover the perceived benefits and shortcomings of group therapy as well as challenges in facilitation from the facilitators' perspective.
In addition, we recruited six participants who had participated in in-person SMARPP programs as clients (referred to as client participants, PC1 -- PC6).
The interviews with them were intended to uncover the experience of group therapy from the clients' perspective, examining the positive and negative episodes in group therapy.
Since there is no fixed group composition for SMARPP, as mentioned in~\ref{sec:smarpp structure}, the facilitator and client participants are not necessarily part of the same group, and thus our results are not limited to a specific group composition.
Additionally, with the exception of two client participants (PC3 and PC5), our participants all had extensive engagement (5 years or longer) in SMARPP and participated in various group compositions, and were thus capable of offering comprehensive comments on SMARPP.
All of the participants had experience with in-person SMARPP, and some of them participated in online SMARPP during COVID-19.
Table~\ref{tab:d_f} and Table~\ref{tab:d_p} show the demographic details of our facilitator participants and client participants, respectively.

\subsection{Interview Structure}
We designed a semi-structured interview to obtain qualitative data about experience in communication between facilitators and clients as well as among clients during in-person SMARPP programs.
The questions used in our semi-structured interview are presented in Table~\ref{tab:faci_q} and Table~\ref{tab:pat_q} in the Appendix.
As an ice-breaker, we first asked participants to describe the general structure and experience of in-person SMARPP programs. 
We then moved to questions that were more directly related to communication practices and experience. 
We specifically asked our facilitator participants questions about what they kept in mind when conducting SMARPP programs and what challenges they perceived, as well as the benefits and shortcomings of group therapy from their perspectives.
To examine client demotivators from the perspective of facilitators, we included questions specifically related to their communication practices that may lead to treatment discontinuation.
We asked our client participants to articulate positive and negative moments regarding communication with facilitators and other clients during SMARPP, aiming to account for motivating and demotivating aspects, respectively.

The interviews generally lasted about one hour.
All the interviews were conducted through an online meeting platform (Zoom) and recorded with the consent of our participants.
We compensated our participants with a gift card of 2,000 JPY (approximately 15 USD at the time when we conducted our study).
All the study participants were Japanese, and we thus conducted interviews in Japanese.
We transcribed the recorded interviews, and performed later analysis in Japanese.
The interviews with facilitator participants were conducted from November to December 2022, and the interviews with client participants were conducted from February to May 2023.

\subsection{Ethical Considerations}
All participants provided informed consent for the study, including the use and analysis of the interview data for research purposes. Furthermore, the interview data collected was analyzed and results were presented in an anonymous manner to ensure participant confidentiality. Recorded interview sessions were deleted upon completion of the study. Ethical approvals for this study were obtained from both the institutional ethics review board at the first author's university and the hospital ethics review board affiliated with the facilitators and clients.

\subsection{Data Analysis}
As the original interviews and analysis were in Japanese, we translated quotes into English for the report presented in this section.
Multiple Japanese-native authors of this paper reviewed the translations to confirm that they were as faithful as possible.
We then performed a thematic analysis on the data we obtained from our interviews using a deductive approach~\cite{braun2006using} to categorize quotes under motivators/demotivators and related themes.
The second author transcribed the recorded interviews, and extracted quotes that related to our research questions.
The first and second authors then jointly conducted the analysis and categorization.
They revised their categorization until both reached a consensus.

\begin{table*}[t]
\small
    \centering
    \caption{Identified motivators and demotivators from facilitators, peers, the structure of the CBT group, and online settings.}
    \label{tab:results}
    \begin{tabular}{p{0.14\textwidth}p{0.38\textwidth}p{0.38\textwidth}}
        \toprule[1pt]\midrule[0.3pt]
         & Motivators & Demotivators \\
        \midrule
        From facilitators
            & 
            \begin{minipage}{0.38\textwidth}
                \begin{itemize}[leftmargin=*]
                    \item Personalized empathy
                \end{itemize} 
            \end{minipage}
            &  
            \begin{minipage}{0.38\textwidth}
                \begin{itemize}[leftmargin=*]
                    \item Power dynamics caused by school-like atmosphere
                \end{itemize}
            \end{minipage}
            \\
        \midrule
        From peers 
            & 
            \begin{minipage}{0.38\textwidth}
                \begin{itemize}[leftmargin=*]
                    \item Positive social influence from peer clients
                    \item Collective sharing and learning
                    \item Positive social influence from peer staff (recovered members)
                \end{itemize}
            \end{minipage}
            & 
            \begin{minipage}{0.38\textwidth}
                \begin{itemize}[leftmargin=*]
                    \item Unequal participation and tension
                    \item Contagious low morale and exposure to triggers
                \end{itemize} 
            \end{minipage}
            \\
        \midrule
        Structural 
            & (Not mentioned)
            & 
            \begin{minipage}{0.38\textwidth}
                \begin{itemize}[leftmargin=*]
                    \item Large group size
                    \item Gender imbalance
                    \item Heterogeneous socialization preferences
                    \item Overly broad diversity
                \end{itemize} 
            \end{minipage}
            \\
        \midrule
        Online 
            & 
            \begin{minipage}{0.38\textwidth}
                \begin{itemize}[leftmargin=*]
                    \item Reduced impact of social anxiety
                    \item Facilitating client comfort and openess
                \end{itemize}
            \end{minipage}
            & 
            \begin{minipage}{0.38\textwidth}
                \begin{itemize}[leftmargin=*]
                    \item Lack of opportunity for outings
                    \item Lack of casual interactions
                \end{itemize}
            \end{minipage}\\
        \bottomrule[1pt]
    \end{tabular}

\end{table*}

\section{Results}

\subsection{Motivator and Demotivators for Engaging Group Therapy related to Facilitators}

    \subsubsection{Motivator: Empathy and Acknowledgement - in a Personalized Way}
    \label{sec: empathy and acknowledgement}

Facilitators play a vital role in motivating clients by demonstrating virtues, such as empathy and acknowledgment, similar to individual CBT. 
Facilitators try to turn SMARPP into a \textit{``place of belonging''} or \textit{``refuge''} (PC4).
Their attempts involve \textit{``expression of empathy''} to participants (PF6), avoidance of \textit{``confrontation and denial, and praise of positive changes''} (PF1).
In the context of substance use, facilitators also \textit{``avoid making value judgments based on social norms''} (PF2).

However, it is sometimes challenging to personalize empathy and attention in practice, particularly in a large group. Although facilitators attempt to treat clients with empathy, they sometimes unintentionally ignore or forget other clients, which can cause negative experiences.
For instance, PC3 told us about a SMARPP gathering that she had joined in the middle of an ongoing session.
She was not given an opportunity to talk about herself during the check-in session and thus was not able to discuss her problems at all. 
However, preventing such incidents is challenging because of logistical constraints.
Although the SMARPP gatherings are scheduled and announced in advance, they accommodate clients arriving late, and there is no definite group composition to give clients the freedom to join or drop out.
PF4 mentioned challenges associated with this particular style: \textit{``Clients change every time, and we take turns as facilitators.
It's not uncommon that it's been more than a month since I last facilitated.
I feel like there were very few people I know.''}

To maintain awareness of the conditions and personalities of each member, PF4 conveyed that after each session, all facilitators employed a collaborative document to record things that caught their attention in their supervised sessions.
However, the shared document does not fully solve the problem, particularly for clients who do not attend SMARPP regularly or who joined recently: \textit{``Besides the spreadsheet, we don't really share [knowledge about clients] elsewhere.
So, if someone doesn't come up in that spreadsheet as someone of interest, they might be overlooked''} (PF4).
PF4 wished to have \textit{``some sort of communication device''} to \textit{``share information with the staff who are located at the back''} about participants' conditions, in case she could not pay attention to everybody.

\subsubsection{Demotivator: Power Dynamics Caused by School-like Atmosphere}
\label{sec: power dynamics school}

Power dynamics between the facilitators and clients can demotivate continuous attendance in group therapy~\cite{bieling2022cognitive}, and our results suggest that group settings resembling a classroom can cultivate such power dynamics.
According to our interviews, facilitators are aware that they should \textit{``avoid creating a hierarchical or authoritative atmosphere as much as possible''} (PF1).
However, our client participants commented that there are still instances where they felt like being treated condescendingly, for instance, like a kindergarten child: \textit{``There's this strangely kindergarten-teacher-like person, and they seem to treat me like I'm stupid or a child''} (PC1).
Our results also suggest that discussion styles led by facilitators may also demotivate participants because of \textit{``a negative association with studying in school''} (PF2). PC5 told us that discussions reminded her of her student era, and this caused a conditioned response of stress: \textit{``In SMARPP, everyone sat in a circle and we were asked to answer what we thought about the topic.
I hate it as it reminds me of my student time when the teacher forced me to answer questions.
It was so stressful for me to think about what to say while waiting for my turn.''}

This perceived clear role separation of ``teacher'' and ``students'' can result in absenteeism or even drop-outs from the entire program. 
 PF2 described how excessive focus on teaching the materials caused clients' dropouts when starting their career as a SMARPP facilitator: \textit{``When I first started this job, I mostly focused on the teaching process using the textbook... and there were many dropouts.''} PF4 expressed similar concerns and was once described by clients as \textit{``teacher-like''}. PC1 expressed dissatisfaction with a facilitator's way of speech, which led to a few-month absence from the therapy sessions: \textit{``So, there was a time when I didn't want to go [because of the kindergarten-teacher-like facilitator]. I took a break for a few months, even though facilitators take shifts every week.''}

While the power dynamics and overemphasis on teaching may be one of the causes of dropouts in group therapy, facilitators can learn to adapt their style to create a more collaborative and engaging learning environment.
For instance, some facilitators chose to avoid too much teaching in group CBT. PF2 discussed that they \textit{``learned to change such a style [overemphasis on teaching]''} and \textit{``avoided making the sessions too academically oriented''}. Accordingly, \textit{``groups became active and the dropouts also decreased''} (PF2).
 In addition, equal conversations such as facilitators' self-disclosure resembling what clients had done could foster the power balance according to PC1: \textit{``It was great when they [facilitators] showed their own weaknesses or shared that they have experienced similar things as us. It creates a sense of unity when there's some disclosure, rather than separating the roles as a facilitator or a client.''}

\subsection{Motivators and Demotivators from Peers}
\label{se: peers}

\subsubsection{Motivator: Positive Social Influence from Peer Clients}
\label{sec:positive influence from peer clients}
Peer clients can stimulate each other's motivation.
PF1 shared that when a client struggling, \textit{`` sometimes a single word from another client is more effective than what facilitators say.''}
Clients are encouraged to share their experiences and triggers in SMARPP groups.
Hearing similar experiences and triggers creates a feeling that \textit{``[one] was not alone and everyone was the same''} (PC2), which led to feelings of emotional resonance when they heard about others' \textit{``hardships in living''} (PC2) and similar worries (PC3).
Reflecting on this, PC3 described SUD as \textit{``a disorder of loneliness''} and shared how staying with peers provided a valuable cure. This effect was also heightened for members of the same minority in the group. For instance, PC3 shared a story of how her presence influenced another female client, which strengthened her own motivation: \textit{``When a woman who was coming for the first time said that she would feel more at ease if I were there, I thought I should go.''}

Our results also suggest that being open to one's situation can be contagious among clients.
PC2 shared her own experience: \textit{``When someone is speaking truthfully when sharing their thoughts, it resonates... Looking at such [honest people], I also thought it was better to become honest, and I felt it was OK to speak more honestly and openly.''}

In the context of CBT, understanding one's emotional and behavioral triggers is an important step in therapy~\cite{beck2020cognitive}.
For people who had difficulties identifying their triggers, hearing from other clients was particularly helpful for recognizing triggers that also applied to themselves: \textit{``As they [clients] gradually listened to more people's stories and progressed through workbooks, they began to discover more and more triggers that were significant to themselves''} (PF1).
For clients who were \textit{``struggling with dealing with emotions,''} PF3 shared that she sometimes \textit{``asked if anyone else has gone through something similar.''}
This helped clients relate to answers from others and thus enhance self-awareness.
PF2 described this phenomenon as \textit{``verbalization,''} \textit{``the process of putting words to experiences and articulating thoughts and feelings.''}
Verbalization through peers was thus an important activity for clients who had difficulties expressing their feelings.
Our results thus revealed that the similarity of clients is one determining factor for the sustained engagement of group therapy.

\subsubsection{Motivator: Collective Sharing and Learning}

Our results suggest that client participants try to embrace people's differences, uncovering new perspectives.
PC6 shared that clients often have difficulty exploring different thoughts and opinions alone because they \textit{``probably thought [only] about themselves and their surroundings''}.
The presence of peer clients thus provides an opportunity to see and think about different perspectives and approaches.
PC1 enjoyed \textit{``exploring the unknowns by listening to others' stories''}.
Thus, obtaining various opinions from other clients also diversified the experience of the SMARPP program.

Our interview participants also shared their perceived merits of being in a group in workbook reading and check-in sessions.
Learning the workbook with others helped PC3 strengthen the retention of knowledge in memory.
Compared to 1-on-1 style therapy, PC6 enjoyed reading the text aloud and engaging in peer learning because of ``the existence of companions''.
Group settings also encourage sharing opinions and experiences as opposed to what is written in the workbook.
PF5 agreed that instead of a smooth group where everybody pretends to be cooperative, challenging what has been taught was more meaningful: \textit{``It's not about getting smarter by memorizing [the textbook]. It's precisely because there's internal resistance [against the textbook] within a person, that sometimes they think, `I'll do it anyway, I'll tackle it.'... I don't think 'smooth' necessarily equals 'good,' you know, it may just appear that everyone is nodding along and seemingly agreeing.''}
PF2 commented on how a group setting can encourage clients to express different opinions: \textit{``In group settings, it's easier to disagree with the textbook.''}
Our results thus revealed group therapy can offer peer learning by encouraging different thoughts, which cannot be achieved in 1-on-1 settings.

\subsubsection{Motivator: Positive Social Influence from Peer Staff}
\label{sec:positive influence from peer staff}
In addition to peer clients, our results uncovered unique value of peer staff (people who recovered from substance use through SMARPP).
In particular, they can function as catalysts for spreading positivity, motivation and honesty, creating a ripple effect of beneficial actions.
PF2 wanted to have more such staff in the group to promote openness and positive connections: \textit{``By building on those initial contributions [from peer staff], other attendees may feel more comfortable sharing their own similar experiences or perspectives.''}

PC4 had experience participating in SMARPP programs as a peer staff member, and commented: \textit{``[clients] would benefit from connecting with certain resources''} and \textit{``guide them to other supports after the program''}, such as external self-support groups.
PC4 valued himself as a bridge among people in groups: \textit{``clients might come because I [PC4] was there.''}

Moreover, peer staff also helped clients in recovery build up their confidence to continue engaging in the program and envision the future.
Clients ended up dropping out of the program due to uncertainties about the future and the treatment efficacy: \textit{``some people have doubts and wonder if this will really work, and there are indeed a certain number of people who drop out because of those doubts''} (PF2).
Recovered peers thus acted as role models and spoke on behalf of struggling clients: \textit{``Having peer staff who can provide such examples can be encouraging and guide what steps to take''} (PF2).
There is often a moment when clients start to feel bored and find no clear progress in the course of the SMARPP program.
PF6 told us that peer staff can be helpful for clients to overcome such moments \textit{``if more experienced peers who have progressed further in their treatment had also experienced such a period, and try to convey that the challenges the person is facing are something everyone can relate to.''}   
Our results thus suggested peer staff's importance in a SMARPP gathering as a role model to spread motivation.

\subsubsection{Demotivator: Unequal Participation and Tension in Group Therapy}
\label{sec: unequal participation}
While our results revealed various benefits of the group setting, we also uncovered demotivating factors associated with it.

Some clients often dominate discussions, potentially evoking frustration among other clients.
PC5 became stressed when monopolizers \textit{``kept talking non-stop''} and worried about the session dragging on.
This can also lead to complaints against the facilitator in charge: \textit{``I wanted the facilitator to stop the person who talks forever''} (PC1).
Our facilitator participants were also aware of this issue, and they attempted to give everyone equal opportunities to speak to create \textit{``a sense of unity''} and that \textit{``nobody is being left out or excluded from the group''} (PF4).
Facilitators used strategies to manage time while still showing a willingness to listen; for example: 
\textit{``I still have some questions I'd like to ask, but I'm a bit worried about the time''} (PF3), or \textit{``Let's talk about it later. I'm here to listen and support you when the session ends''} (PF4).

Discrimination against minorities can also contribute to group tension.
PF2 acknowledged that the presence of ethnic or sexual minority individuals within the group could lead to the formation of subgroups and the potential for further discrimination within these \textit{``minorities within minorities.''}
PC4 expressed his discouragement over constantly disclosing his sexual orientation because of changing group members: \textit{``I wonder how many times I should mention it, especially when the group dynamics keep changing.... It's a bit disheartening to come out every time.''}
In addition, discrimination could occur between those who used legal and illegal substances.
For instance, PC1 expressed their discomfort when another client said things like \textit{``Don't lump me with them [people who use methamphetamine]. We are not doing anything illegal''}, which was also echoed by PC3.

Tension could also arise from specific topics discussed within the group.
PF8 mentioned that topics such as debates on marijuana prohibition can generate tension among clients.
Moreover, while sharing triggers is considered beneficial in general, our interview participants mentioned that it can negatively impact clients. PF7, for instance, stressed the importance of avoiding descriptions of triggers in too much detail as it could also stimulate others' cravings.

However, intervening with clients for this reason could also lead to frustration when they simply wanted to disclose relevant personal experiences (PC1): \textit{``A member who got angry [as stopped by a facilitator when disclosing episodes triggering others' cravings] was like, 'I can't even say this?'.... It's tough, we don't know who will be triggered by what trivial things.''}
These results suggest the challenge of achieving the right balance between the freedom to share information and its potential influence.

\subsubsection{Demotivator: Contagious Low Morale and Potential Exposure to Triggers in Group Therapy}
\label{sec: contagious low morale and potential exposure to triggers}
In some situations, there were clients in a group who showed little or no motivation to quit substance use, or even tried to sell drugs to others.
The presence of such clients may degrade the experience of SMARPP and create a contagious low morale in the group.
As PC1 commented: \textit{``There were times of good waves and times of bad waves. In a bad wave, people exchange information about drugs and drug dealers and relapse together.''} 
PC6 shared his frustrating experience of seeing people \textit{``who confidently asserted that they have no intention of quitting''} and \textit{``who said they would casually use drugs.''}
PF1 shared several examples of negative comments that can demotivate other clients: \textit{``It's meaningless to do this,''} \textit{``Talking about [substance use] here only increases desires,''} and \textit{``There are probably a lot of drug dealers here.''}
PC2 mentioned how such negative peer clients could cause her to \textit{``[feel] a bit annoyed, and at the same time envious [about using drugs].''}
Others' sudden dropouts can also create negative feelings in a group as PC1 said: \textit{``I wonder what happened when someone stopped coming for a while. I sometimes imagine if they got arrested or even died. Or they're doing well but just got tired of SMARPP.''}
PC4 also encountered clients selling drugs to him after the sessions. Although he did not buy them, he had a sense of regret: \textit{``I felt a bit disappointed, and I felt like I appeared weak to others at the moment.''}

Facilitator participants recognized these issues and attempted to resolve them by maintaining positive attitudes and acknowledging clients success, and by controlling discussions when they became potential threats or risks to other peer clients during sessions: \textit{''If the person's story seems to touch on something that may compromise safety or feels difficult to handle within the group, we address it individually later''} (PF1).
PF7 told us that while facilitators clearly prohibited clients from inviting drug usage or selling drugs, she admitted that reinforcing the rules could have a negative impact on group dynamics: \textit{``If we become too focused on enforcing strict rules within the group, the treatment environment can become very formal and rigid. It may even lead [clients to form] a sense of hostility or antagonism against us.''}

However, outside of the sessions, the group nature of SMARPP allows clients to connect directly, potentially enhancing negative behavior.
PC4 stated: \textit{``Discussions involving drug use or those with a focus on pleasurable desires tend to get lively outside of SMARPP. It inevitably shifts the conversation in that direction, and I find myself trying hard not to get anxious or worked up about it''} (PC4).
PC1 added: \textit{``There were multiple times a male and a female client developed a very close relationship and dropped out together.''}

PF1 stated the groups' therapeutic outcomes depended on the composition of clients and their closeness: \textit{``When a group consists mostly of individuals who are at high risk, there is a greater likelihood of everyone relapsing together due to the close relationships formed within the group. On the other hand, when the group consists of highly motivated individuals, maintaining those lateral connections can also contribute to positive outcomes''}(PF1).

\subsection{Structural Demotivators}

\subsubsection{Demotivator: Overly Large Group Size}
\label{sec: group size}

A typical SMARPP group has 10--15 people, while sometimes a large group can have 20--30 people (PF1, PF2, PF3, PF7, PC1). When the group size becomes too large, it can create a sense of distance and hinder meaningful interaction. 
PF3 commented: \textit{``In a large room, it's difficult to intervene or get involved when observing from a distance.''}
PF1 acknowledged a similar issue, but noted that in a larger group, the impact of participants who bring negative influence may become smaller: \textit{``As the number of people increased significantly, there seems to be a natural decrease in individuals who disrupt the group excessively.''}

\subsubsection{Demotivator: Gender Imbalance}
\label{sec: gender imbalance}

Gender balance can also impact group dynamics.
According to PC1, a typical ratio of gender in a SMARPP group is 9:1.
Female clients are typically under-represented, potentially creating a less welcoming atmosphere for them and other gender/sexual minorities (PC3).
For example, PF6 shared his opinion that it was particularly difficult for teenage girls to speak up in a group where adult males are the majority.

\subsubsection{Demotivator: Heterogeneous Socialization Preferences}
\label{sec: social prefs}
Differences in socializing preferences among clients posed another demotivating factor.
PF3 told us: \textit{``There are some people who are inclined to get closer and be friendly while others prefer to maintain a certain distance due to their inherent difficulty in socializing.''} 
Our interview participants suggested that preparing different group sizes could mitigate this issue.
During the COVID-19 pandemic, the hospital divided clients into a large and small group when the number of clients were over the threshold for one group.
In such a situation, clients were able to \textit{``choose the venue based on their preferences''} (PF8).
For example, individuals who preferred not to draw much attention would join a big group while others who could be less nervous in front of fewer people and who wanted to share rather private stories could go to the smaller group.
However, such flexible group formations would not be always possible due to the limited number of SMARPP facilitators: \textit{``If we had enough manpower, we could create more specific group divisions, allowing us to provide support tailored to each individual's needs''} (PF6).

\subsubsection{Demotivator: Overly Broad Diversity}
\label{sec: substance diversity}

While diversity allows for gaining insights from different perspectives, it can also create challenges when empathizing with others.
According to PF6: \textit{``Currently within SMARPP, there are people using different substances like amphetamines, marijuana, over-the-counter drugs, and prescription medications mixed together. Additionally, the age range is quite diverse, spanning from teenagers in their 10s to individuals in their 50s, including both males and females....it can be challenging to empathize with others in that diverse group''} (PF6).
PF4 mentioned there were also cases where clients were rejected by their peer clients in SMARPP due to different substance usage: \textit{``Alcohol-only clients probably wouldn't understand much about other substances. Being exposed to various substances, they could even have a risk of using drugs.''}
Thus, group therapy may suffer from weakened empathetic interactions due to diverse client backgrounds and conditions.

\subsection{Motivators and Demotivators of Online SMARPP}
    \label{sec: online settings}

During the COVID-19 pandemic, group therapy in SMARPP was often held online via meeting tools. 
Besides enhanced accessibility, our interviews found that this online format had both motivational and demotivational factors.

\subsubsection{Motivator: Reduced Impact of Social Anxiety}
\label{sec:social anxiety}
Online SMARPP can accommodate clients with social anxiety better than in an in-person setting.
PF7 agreed that the online setting benefits people with social anxiety and can encourage their self-disclosure: \textit{``Those who tend to get nervous in public situations may not feel as anxious [online]. There are times when people can honestly disclose themselves if online''} (PF7). This idea was echoed by PC3, as she once felt strong anxiety as gender minority in the group.
Among our interview participants, PC5 showed her preference for online SMARPP compared to in-person sessions. PC5 had social anxiety around talking in front of the group in person and this was one of the main reasons she dropped out of the in-person program after one cycle. Though in her case, going out for in-person SMARPP was painful because one of her triggers was \textit{``going out and seeing places where she had drug transactions.''} In addition, PC5 could feel a sense of presence even online: \textit{``Because even online, if you show your face, it can feel like you're meeting with everyone, don't you think?''} 

\subsubsection{Motivator: Facilitating Client Comfort and Openness}
Our results suggest that online settings can facilitate comfort and openness.
PC3 commented that an online setting can create a space where individuals feel more comfortable engaging in candid conversations: \textit{``In face-to-face interactions, there are still people who can't be fully trusted, or there are so many people around, so it's difficult to speak up. However, when it comes to online communication, it's easier to be straightforward.''}
PC2 also agreed that without the full presence of physical appearance by turning off the cameras, there is less fear of judgment or exposure, allowing participants to share openly with comfort.

\subsubsection{Demotivator: Lack of Opportunity for Outings}
In-person SMARPP provides opportunities for exercising and traveling not offered by the online setting. PC3 appreciated in-person SMARPP providing her with opportunities to go out and exercise. PF5 also mentioned taking in-person SMARPP helps participants stay clean and tidy. Traveling can offer extra reflection time for the discussion in sessions (PF6) and time away from drugs (PF8). It can also bring \textit{``a sense of accomplishment that comes from investing time and effort to be present physically''} (PF8).

\subsubsection{Demotivator: Lack of Casual Interaction Online}
In-person settings enable nonverbal cues and casual interactions, which are all missing in online SMARPP. The institution where the SMARPP program takes place allows participants to turn off cameras and microphones during online meetings in Zoom-like settings. PF4, PF7, and PC6 expressed their concerns about being unable to see everybody's body language or facial expressions online. For instance, PC3 lamented that she could not see facial expressions or hear laughter online that creates \textit{``a sense of connection with other people''}. PF8 also mentioned participants lost the opportunities to \textit{``engage in small interactions while sitting next to each other''}. PC2 additionally commented without nonverbal cues, \textit{``even if people talk about things like made-up stories, it's not likely to get exposed.''}

PC2 suggested conducting online sessions at the beginning for newcomers to reduce distress related to showing their faces and talking in front of strangers, and then gradually transitioning to in-person settings: \textit{``For someone who is contemplating whether or not to participate with hesitation, the barriers [of online SMARPP] would likely be low... After participating online several times, the desire to share the physical space with others may arise.''}
PF4 also pointed out the possibility of telepresence: \textit{``We could have a robot in the form of a human... not just a screen projection, but an actual physical presence.''}

\subsection{Communication Among Stakeholders at Different Phases in Group Therapy}
\label{se: communications}

Our study clarifies motivators and demotivators for facilitators, peer clients, group structure, and environments in group therapy (RQ1). Next, we illustrate our findings about communication among stakeholders in different phases of group therapy to answer RQ2. 

\subsubsection{Before sessions}
\label{se: before session}

\paragraph{Facilitator to facilitator:} 
Before each CBT session, communication takes place among different facilitators to share information about the sessions they have managed.
Facilitators work in shifts to lead groups with diverse compositions, making effective communication among them essential to ensure continuity and coherence in the therapeutic and group process.
Major motivating factors for clients obtained from facilitators are \textit{empathy, acknowledgment, and personalized attention}.
To achieve them, pre-session communication among facilitators is vital as it allows facilitators to gather information about each participant's experiences in previous sessions, including those led by other facilitators, to be addressed in the following sessions.

However, as we found in Section~\ref{sec: empathy and acknowledgement}, maintaining effective communication between facilitators of group therapy is challenging, both because of the facilitation schedule and the changing roster of clients. Our facilitator participants made use of technological solutions (i.e. shared documents) to aid this process, but our results indicate that the unique challenges associated with this form of therapy could benefit from more targeted tools.

\subsubsection{During sessions}
\label{se: during session}
\paragraph{Facilitator to client}
During the session, communication occurs among all group attendees. 
our results in Section~\ref{sec: unequal participation} suggest that facilitators' ability to ensure equal participation of every attendee is crucial for constructing a positive atmosphere and enabling effective communication during group therapy sessions.  
However, our results in Section~\ref{sec: power dynamics school} demonstrate that other factors, such as \textit{adopting a classroom-style psychoeducation approach}, may also have a substantial effect on clients' sense of power imbalance if not well managed. This suggests that supporting the facilitator's awareness of both power dynamics and the ability to tailor the balance of session content to a specific group's needs could substantially enhance treatment success.

\paragraph{Client to client:} 

As we found in Section~\ref{se: peers}, communication between clients is one of the defining benefits of group therapy 
because it can provide clients suffering from a variety of mental health issues with social support, de-stigmatization, and vicarious learning. 
A variety of factors, including unequal participation (Section~\ref{sec: unequal participation}), heterogeneous socialization preferences (Section~\ref{sec: social prefs}), gender imbalance (Section~\ref{sec: gender imbalance}), discrimination against minority group members (Section~\ref{sec: unequal participation}), heterogeneous experience with substance use (Section~\ref{sec: substance diversity}), and even group size (Section~\ref{sec: group size}) may cause tension and dominance issues within treatment groups.

\subsubsection{After sessions/Outside sessions}
\label{se: after session}
\paragraph{Client to client}

Post-session communication between clients offers potential for both risks and benefits.
Our findings reveal that post-session communication between clients can lead to contagious low morale and exposure to potential triggers, as described in Section~\ref{sec: contagious low morale and potential exposure to triggers}.
One noteworthy aspect of post-session communication among clients is the absence of established norms and facilitation, which can cause conversations to go beyond boundaries and become unsafe.
Forming overly close relationships among clients can also increase the risk of collective relapse and dropouts.
Although these risks may be most associated with group therapy for addiction and associated mental illnesses, our results suggest the need for a nuanced approach to facilitating social networks between clients, as simply connecting clients to each other may have the potential to harm, rather than enhance, treatment outcomes.

\section{Discussion}
While previous research within HCI has predominantly focused on strategies for fulfilling the three basic psychological needs proposed by Self-Determination Theory (SDT) — namely \textit{Competence}, \textit{Autonomy}, and \textit{Relatedness}, our present study suggests that complexities arise in simultaneously satisfying all three needs in the context of group therapy. 
Our findings highlight that conflicts and trade-offs emerge in meeting these basic needs, with tensions manifesting both intrapersonally and interpersonally.

\label{sec: sdt conflicts}

\subsection{Intrapersonal Conflicts of Motivation} 
Conflicts manifest in intrapersonal needs between the desire for \textit{Autonomy} and the pursuit of \textit{Competence} illustrated in Fig.~\ref{fig: tensions} (a). We observed trade-offs faced by facilitators in providing informational support to clients during sessions. Some facilitators expressed concerns that emphasizing the teaching of CBT skills might potentially undermine client motivation, although such skills are important for fostering client \textit{Competence}. Clients, in turn, noted instances where they perceived unbalanced power dynamics with facilitators during instructional sessions, reporting a sense of diminished \textit{Autonomy} even when facilitators conscientiously sought to mitigate such dynamics. Drawing on the theory of power influence~\cite{french1959bases,raven1964social}, it becomes evident that informational or expert power is at play, as clients inevitably view facilitators as ``experts in treating mental health issues'' and as individuals possessing extensive knowledge about addressing their concerns when knowledge is imparted. This phenomenon is further accentuated by recollections of a ``school-like atmosphere,'' where strong power dynamics were experienced with teachers during clients' formative years, regardless of the facilitator's intended approach. Consequently, conflicts between clients' basic needs for \textit{Competence} and \textit{Autonomy} may arise within group therapy.

Another identified potential tension lies in the interplay between clients' \textit{Relatedness} and \textit{Autonomy} illustrated in Fig.~\ref{fig: tensions} (b). Our findings underscore clients' expressed desire to adapt their socialization preferences and the freedom to engage in casual interactions beyond formal sessions. However, negative consequences can arise from such actions, especially when one or both parties have low motivation. The formation of close relationships can then be accompanied by a loss of motivation, leading to collective relapses and the disconnection from positive influences after clients' dropouts. Additionally, structural factors, such as the choice between online and onsite formats, reveal conflicts between clients' basic needs for \textit{Relatedness} and \textit{Autonomy}. While the online format affords clients the \textit{Autonomy} of participating from their homes, it entails the sacrifice of opportunities for social cues from fellow group members. Furthermore, diverse group compositions, including variations in gender or clients' conditions, offer opportunities for exploration of the unknown, thereby enhancing intrinsic motivation to participate. However, this diversity may simultaneously limit opportunities for obtaining empathy from those with similar experiences.

\subsection{Interpersonal Conflicts of Motivation}

The group therapy format introduces challenges where diverse clients may experience conflicts in meeting their basic psychological needs. 

\begin{figure*}[t]
    \centering
    \includegraphics[width=\textwidth]{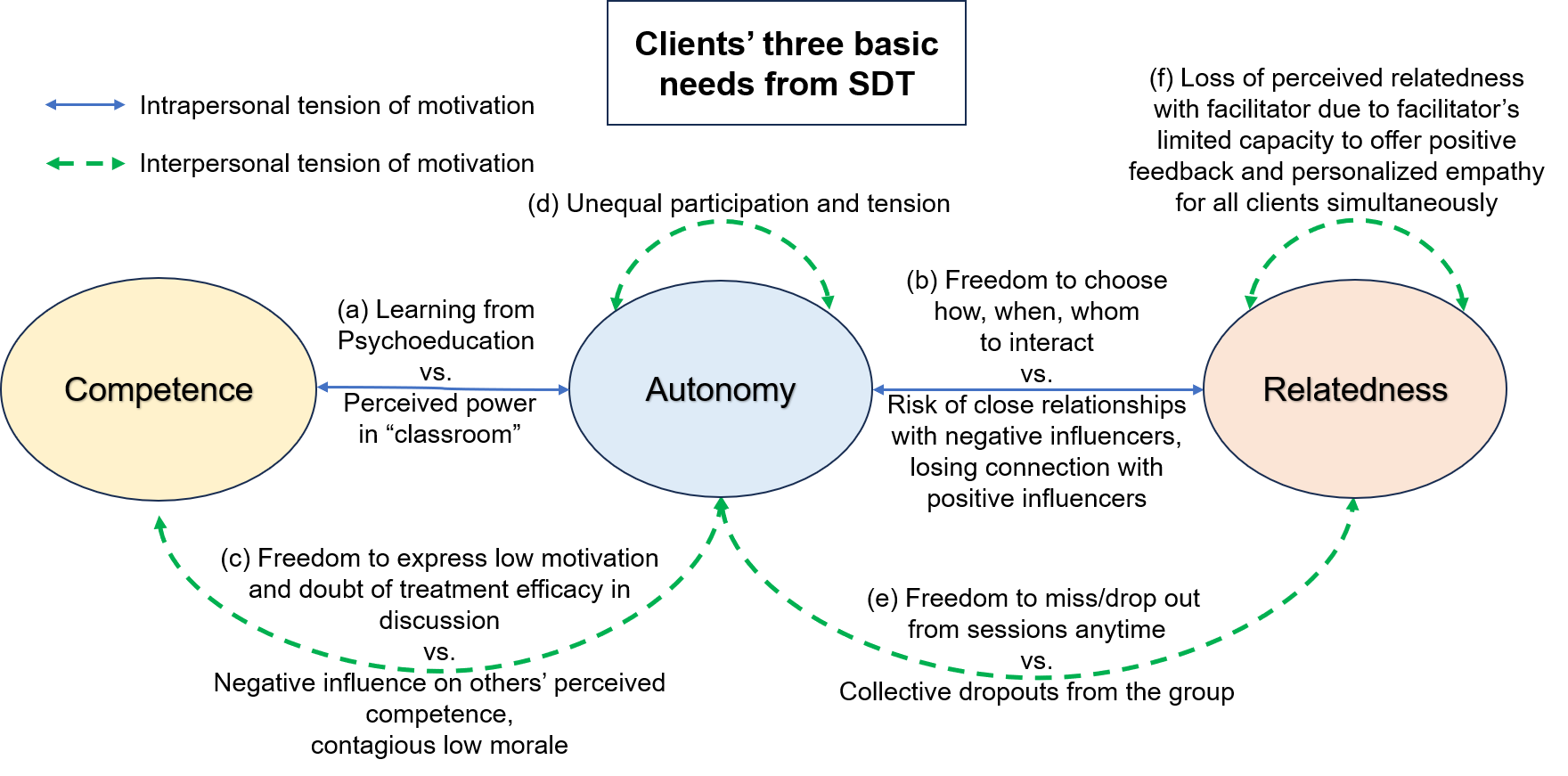}
    \caption{Intrapersonal and interpersonal tension of motivation through SDT in the context of group therapy.}
    \label{fig: tensions}
\end{figure*}

In group therapy, facilitators play a key role in providing positive feedback to clients. According to cognitive evaluation theory~\cite{deci2013intrinsic}, a sub-theory of SDT, such positive feedback from facilitators enhances the intrinsic motivation of clients to engage in recovery because it strengthens their internal locus of causality. Our research findings highlight the significance of delivering personalized empathy in the group setting by providing such positive feedback. From the facilitator's standpoint, concurrently addressing multiple individuals may result in varying levels of positive feedback and personalized empathy offered to each client. The inherent limitations in the facilitator's capacity and memory to effectively manage and respond to personal details from different clients may create tensions in fulfilling each client's basic psychological needs of \textit{Relatedness} with the facilitator, as illustrated in Fig.~\ref{fig: tensions} (f). 
While prioritizing the recollection of specific details about one client's experience, the facilitator may inadvertently overlook another's.
Yet from the client's perspective, the facilitator is perhaps the most discerning person in the group to provide empathy and affirm their small efforts toward recovery. When such expectations are unmet, 
clients may experience substantial dissatisfaction. Our findings underscore the significance of personalized empathy in fostering a sense of \textit{Relatedness} with the facilitator and consequently enhancing motivation to engage in group therapy, as well as the conflicts that emerge among clients with regard to satisfying their need for \textit{Relatedness} with the facilitator.

Moreover, unequal participation and potential tensions during discussions may impede certain clients' motivation to actively engage in group discussion—an integral aspect of group therapy. Despite the essential nature of expressing one's ideas to fulfill \textit{Autonomy} in group discussions, this may lead to the extended monopolization of discussions and offense to others, thus compromising the \textit{Autonomy} of other clients. 
We also observed that clients in group therapy can be doubtful about their \textit{Competence} in recovery due to influence from peers' relapses and low motivation.
Similarly, the freedom to miss sessions or drop out from group therapy at any time poses a risk. This sometimes causes concern among the remaining clients, and can lead to negative social influence and collective dropouts. Consequently, the satisfaction of one client's \textit{Autonomy} may inadvertently compromise the \textit{Competence}, \textit{Autonomy}, and \textit{Relatedness} of others within the group, as illustrated in Fig.~\ref{fig: tensions} (c), (d) and (e), respectively.

\subsection{Implications for HCI: Designs for Balanced Motivational Dynamics}

Our findings regarding motivators and demotivators align closely with existing literature on group therapy practices.
Specifically, the unique benefits of group therapy, as identified by Yalom's small-group factors, are corroborated by our results.
Yalom's factors highlight that group therapy can instill treatment hope, as discussed in Section~\ref{sec:positive influence from peer staff}, as well as provide a sense of not being alone and facilitate information sharing and openness, aligning with Section~\ref{sec:positive influence from peer clients}.
Our results further demonstrate that clients are motivated by personalized empathy from facilitators.

Regarding demotivators, prior work documents several challenges that facilitators face in managing group processes, particularly those arising from specific client characteristics, such as overbearing clients who dominate discussions, disbelievers who challenge treatment outcomes, and drifters who frequently miss sessions~\cite{bieling2022cognitive,yalom2020theory}.
Our results in Section~\ref{sec: unequal participation} and Section~\ref{sec: contagious low morale and potential exposure to triggers} align with these challenges.
However, unlike previous work, our findings suggest that these challenges may stem more from conflicting client needs, rather than solely from client characteristics.

Facilitators can make psychoeducation interactive by engaging clients with questions to avoid a classroom-like atmosphere~\cite{bieling2022cognitive}. Nonetheless, as discussed in Section~\ref{sec: power dynamics school}, we found that clients still experienced teacher-student power dynamics even in interactive group discussions, which reminded clients of their pasts as students, thus suggesting that expert power dynamics may still persist. 

Finally, our findings on structural demotivators, such as inappropriate group sizes or levels of diversity, also align with existing literature~\cite{bieling2022cognitive,yalom2020theory}.
Our work further demonstrates the influence of online group therapy on client motivation, including reduction of social anxiety and promotion of openness, while also highlighting common challenges associated with online meetings, such as the absence of casual interaction due to insufficient social cues being exchanged online~\cite{cristea2019get, yang2022distance}.
Moreover, Bieling et al. suggested having a carefully selected and fixed group to establish group cohesiveness~\cite{bieling2022cognitive}.
However, our study indicates that it is challenging to select clients and develop a fixed group in practice because it can violate client autonomy and requires significant effort from facilitators, who are already occupied with managing many clients and addressing other issues.

\begin{figure*}[t]
    \centering
    \includegraphics[width=0.8\textwidth]{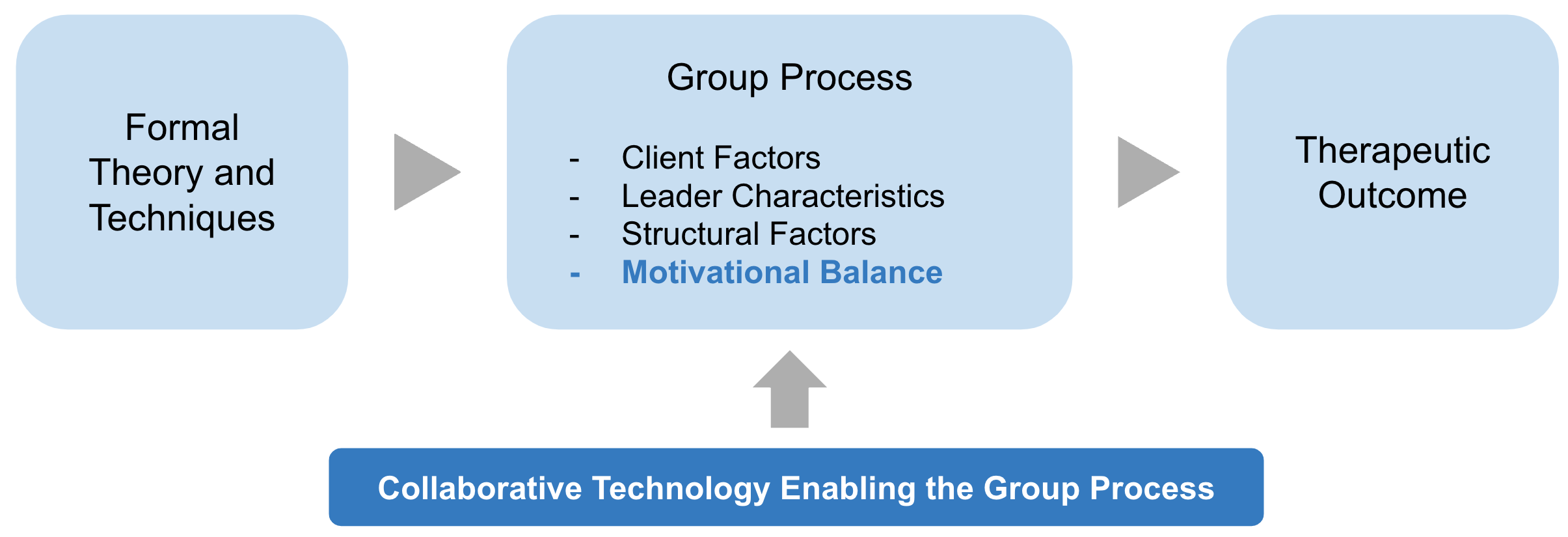}
    \caption{Our augmented model based on Burlingame et al.'s group model~\cite{burlingame2004small,bieling2022cognitive,burlingame2013change} for computer-mediated group therapy In addition to the client, facilitator, and structural characteristics, we introduce a new component, motivational balance, to the group process. We highlight the potential for leveraging collaborative technology to enable the group process.}
    \label{fig: model}
\end{figure*}

Based on Burlingame et al.'s group model~\cite{burlingame2004small,bieling2022cognitive,burlingame2013change}, therapeutic outcomes are influenced by formal theories, therapeutic techniques, the dynamics of group processes including client personal and interpersonal characteristics, facilitator leadership qualities, and structural aspects (e.g., session length and frequency), as shown in Fig~\ref{fig: model}.
Although our findings align with this model in the sense that client motivations are influenced by peer, facilitator, and structural factors, they also indicate that although facilitators may possess effective leadership qualities, they face limitations in addressing specific demotivating factors.
Moreover, client outcomes are significantly influenced by conflicting needs rather than solely by their inherent characteristics. 
Given the complexity of these factors, we argue that beyond theory and techniques from psychology, there exists significant potential for collaboration between mental health and HCI researchers to enhance treatment outcomes of group therapy through technological interventions.
Thus, we propose an augmented model for computer-mediated group therapy as shown in Fig.~\ref{fig: model}. 
In addition to the formal theory and techniques outlined in the original model, we introduce a new component, motivational balance, to the group process. We also highlight the potential for leveraging collaborative technology to enable the group process.

While prior HCI research on motivation primarily focused on individual motivation, our contribution extends to the collective motivation and motivational dynamics within a group. 
Regarding supporting individual motivation, Hamid et al. developed a storytelling game to deliver CBT content to college students to promote intrinsic treatment motivation~\cite{hamid2022you}.
Stawarz et al. designed a platform to enhance relatedness between the therapist and client for individual therapy~\cite{stawarz2020integrating}.
Although some HCI research has shown that integrating client social networking can enhance individual treatment motivation~\cite{doherty2012engagement,lederman2014moderated}, integrating collective learning among peers in these systems requires careful design considerations, as our findings suggest that potentially unintended outcomes may arise, leading to motivational conflicts.
Therefore, our work advocates for future research to develop collaborative technology adapted to the group therapy environment, enhancing collective motivation for group treatment, in addition to the existing technologies supporting individual therapy.

\section{Research Opportunities}
\label{sec: research opportunities}
In the subsequent sections, we demonstrate four research directions for the HCI community to address and alleviate intrapersonal and interpersonal tensions related to basic needs and motivation within group therapy contexts.
This involves supporting facilitators' record keeping when addressing clients' interpersonal conflicts in \textit{Relatedness} with the facilitator (i.e. (f) in Fig.~\ref{fig: tensions}; providing training for facilitators in handling intrapersonal conflicts about \textit{Competence} and \textit{Autonomy}, and interpersonal conflicts about \textit{Autonomy} (i.e. (a),(d) in Fig.~\ref{fig: tensions}; assisting clients in establishing social boundaries when confronted with conflicting needs of \textit{Autonomy} and \textit{Relatedness} (i.e. (b) in Fig.~\ref{fig: tensions}; and constructing virtual group therapy platforms utilizing multiple agents that possess awareness of the interpersonal tensions in basic needs (i.e. (c),(d),(e) in Fig.~\ref{fig: tensions}).

\subsection{Supporting \textit{Facilitators'} Record Keeping}
\label{sec: implication for facilitators a}

Providing personalized empathy and attention to each client is one of the primary responsibilities of facilitators both according to our results and prior work~\cite{bieling2022cognitive}, and helps mitigate the potential power dynamics inherent in group settings that trigger client interpersonal conflicts in \textit{Autonomy} and \textit{Competence}.
However, because facilitators and client group composition are in constant flux, record keeping and information sharing between facilitators is critical to enable facilitators to have the necessary context to perform this function. 

There are existing digital systems that support clients in recording their personal health data for sharing with clinicians, but previous studies have identified that these records often contain extensive information, posing challenges for therapists to review efficiently in individual treatment~\cite{liu2011barriers}.
It can be more challenging in group contexts, where facilitators need to navigate through collective client data. Moreover, this task is complicated by several factors unique to the group therapy context. 
As described in Section~\ref{sec: empathy and acknowledgement}, many facilitators may take turns working with the same group, meaning that information must be recorded and stored in ways that will allow all facilitators to gain sufficient understanding of a potentially large number of past sessions at which they were not present. 
This suggests that the notes cannot serve as simple memory prompts, but must be fully informative in and of themselves. 
Furthermore, due to the continuous changes in the group composition, it is possible that not all group members will be mentioned adequately. This is because individual facilitators have to be selective in what they record.
Murray et al. developed a system for clinicians in the emergency department to manage clinical documentation for up to 35 patients at a time, yet they use data from Electronic Health Records (EHRs) instead of retrieving information from ongoing meetings. 
Thus, supporting recording, recalling, and sharing of session information for facilitators in group therapy has the potential to be a fruitful future research direction in HCI.

Digital support tools may thus have the potential to ameliorate some of these challenges if tailored to the specific needs of group therapy facilitators. Meeting support tools have a long history in HCI~\cite{streitz1994dolphin, waibel2003smart, wolf1992communication}, and have addressed both the need to record and recall information generated during meetings. 
More recent work has used natural language processing techniques for generating automatic meeting transcripts and summaries~\cite{murray2008summarizing,song2021smartmeeting}. 
However, to the best of our knowledge, there has been no work investigating how such technologies could be adapted to benefit group therapy, where the group composition is constantly changing.
Yet, we also note that privacy concerns will be paramount in the case of mental health treatment, particularly given existing concerns about the privacy of other digital mental health support tools~\cite{parker2019private}.
Future research should, therefore, be aware of and consider the privacy limitations when supporting meetings in group therapy.

\subsection{Supporting \textit{Facilitators'} Skill Training}
\label{sec: implication for facilitators b}
In the current structure of group therapy, facilitators are also the primary source to intervene and mediate during moments of interpersonal tension.
Our results uncovered challenges facilitators faced until they had enough experience.
For instance, some clients may monopolize the session or create tension in the group due to conflicting opinions (Section~\ref{sec: unequal participation}), while some clients with social anxiety may require special treatment (Section~\ref{sec:social anxiety}). Our facilitator participants also found it challenging to balance the amount of time for psychoeducation in a session and reduce induced power dynamics (Section~\ref{sec: power dynamics school}).
Failure to handle these situations often discourages client engagement in group therapy.

This points to a research opportunity for further study to support the training of group therapy facilitators. For example, previous research in HCI has explored communication training systems using virtual agents, such as the systems developed by Hoque et al. for social skills training in interviews or remote discussions~\cite{hoque2013mach,samrose2022mia}. 

However, these systems have been designed for 1-on-1 settings. Our research, conducted in a group therapy context, suggests that future research could focus on designing systems that can effectively support facilitation training in group therapy using multiple virtual agents. Such a system would simulate an in-person group therapy session, with multiple virtual agents playing the roles of different types of clients. The composition of virtual clients would be adjusted depending on the experience of the trainee. For example, novice trainees would be provided with a set of friendly clients, while more experienced trainees would be presented with more complex scenarios, such as clients who exhibit an offensive attitude towards the facilitator.
Nevertheless, the realization of such innovations may raise ethical concerns regarding the intentional incorporation of offensive, challenging, or even biased behaviors in virtual agents for facilitator skill training, that may oppose the concept of responsible AI.
Future work should establish design guidelines for such systems so that HCI research can contribute to designing effective and ethical training systems.

\subsection{Supporting \textit{Clients'} Balanced Social Boundaries}

\label{sec: implication for clients a}
Clients in group therapy may encounter difficulties in establishing appropriate social boundaries with their fellow members, as described in Section~\ref{sec: contagious low morale and potential exposure to triggers}.
Previous studies suggest that closer social distance can have positive effects on treatment outcomes, such as promoting collaborative learning~\cite{khodayarifard2010effects,lewinsohn1999psychosocial} and building empathetic connections with peers~\cite{barrowclough2006group,barrera2013meta,bramham2009evaluation}, factors particularly relevant in the context of group therapy~\cite{bieling2022cognitive}. 
Existing research in HCI for mental health therapy often focuses on digitalization or gamification for individual clients~\cite{andersson2014internet,takano2016web,hamid2022you} while there is a notable lack of studies on platforms that support client communication in group therapy environments.
Some studies have proposed methods to empower patient communities, such as embedded social networks enabling client interactions~\cite{doherty2012engagement,lederman2014moderated}.
However, research on how such platforms can be integrated into formal therapy programs for positive peer influence remains limited.
Future research should thus explore integrating a client community to promote positive connections and establish relationships among group members in group therapy.

Our findings also highlight that instances of heightened social intimacy among group members, particularly in cases where close bonds are formed between newcomers or those who possess lower motivation, can result in unfavorable treatment outcomes and even trigger relapses into drug misuse.
Prior research has shown that other mental health issues (e.g., depression) may also spread among social networks according to the contagion theory~\cite{bastiampillai2013depression}. This suggests that our findings about the contagious nature of reduced morale in group therapy are relevant to other contexts as well, although further research is naturally warranted.

Current strategies employed by facilitators in our findings to tackle the challenge of social boundaries may hinder the effectiveness of treatment by preventing the formation of emotional connections among clients and creating a controlling environment that limits clients' \textit{Autonomy} to interact with each other. Therefore, it is crucial to assist clients in finding balanced social boundaries.

While some prior work has discussed social boundaries in the context of social networks~\cite{shi2013using,stutzman2012boundary}, little attention has been paid to developing technology that helps clients undergoing mental health treatment establish appropriate social boundaries, and as a result, there is potential for future efforts to design tools that empower individuals to create and maintain suitable social boundaries. 

For example, researchers could explore how clients communicate and exchange information with each other outside of formal therapy sessions in group therapy. Additionally, there are opportunities for research and development of technology that supports client communication and maintains balanced social boundaries after sessions while respecting their \textit{Autonomy} to interact with each other. For instance, nudges can be a potential mechanism to aid clients in regulating their social boundaries during online communication. An illustrative example is the work of Masaki et al., who created a nudge system to assist adolescent social media users in avoiding privacy and safety threats~\cite{masaki2020exploring}, which can be further extended to address issues in group therapy. For instance, machine learning algorithms could be developed for automatic detection of ``boundary-crossing'' content, and explore different nudge designs that can effectively help clients stay at a safe social distance.
However, collecting training data for developing such models requires careful consideration of individuals' privacy.
Additionally, it is not yet clear whether gentle nudges could also cause reactance in clients, potentially affecting their autonomy to interact with each other.

\subsection{Facilitating Balanced Peer Support with Multiple Agents for \textit{Clients}}

\label{sec: implication for client b}

Our findings highlight the benefits of incorporating peers into group therapy. 
In group therapy settings, clients experience a blend of advantages derived from both facilitators and peers. 
However, our results also highlight the challenges associated with real peers in group therapy, as varying social preferences and motivational conflicts among clients and their peers can arise. The selection of suitable peers requires meticulous coordination, which may not always be feasible. 
Additionally, privacy concerns emerge when real-person peers are involved, as clients may share sensitive or intimate information during CBT sessions. 
These privacy threats pose a significant deterrent to engaging in group therapy~\cite{lasky2006confidentiality}.

These findings point to a possible research direction leveraging chatbots to find balanced peer support in group therapy. 
Earlier research has demonstrated that users can establish relationships and attachments with chatbots, mirroring the dynamics observed in human-human relationships~\cite{xie2022attachment,skjuve2022longitudinal,skjuve2021my}. 
Additionally, the effectiveness of chatbots in addressing mental health issues has been well-explored~\cite{vaidyam2019chatbots}. 
Notably, the phenomenon of chatbot self-disclosure of mental issues has been found to elicit a reciprocal effect on users~\cite{lee2020hear}, similar to our observations that clients experience positive social influence from peers in group therapy.

However, prior work has focused on interaction between one bot and one user, while our findings have demonstrated the distinct advantages of group learning and collective sharing in group therapy. Although prior work has shown that mental health chatbots in 1-on-1 interactions can provide various types of social support, such as informational and emotional support~\cite{bae2021social}, integrating multiple agents to simulate a group therapy environment could potentially offer the unique advantages mentioned in Section~\ref{sec:positive influence from peer clients} and Section~\ref{sec:positive influence from peer staff}.
Future work in HCI could thus explore the design of virtual group therapy environments employing multiple virtual agents. For instance, these environments would include one real user as a client and multiple virtual agents as either the facilitator or peers. 
It is worth noting that we do not intend to imply that virtual agents or chatbots can replace genuine human connections, but they have the potential to provide peer support in instances when preferred human interactions are not possible.

As peers in this context are chatbots, the system can easily control their profiles to align with the real client's concerns and interests. Privacy concerns can also be reduced, as the system involves no real person besides the client. This also eliminates the interpersonal conflicts in motivation we discussed, as the user takes precedence over the virtual agents. But it should also be noted that as this remains an emerging field, sensitivity to the ethical issues and dilemmas related to chatbots in mental health treatment is of the utmost importance for researchers and developers~\cite{coghlan2023chat, cabrera2023ethical, vilaza2021automation}. Particularly, in simulating peer behaviors for a group therapy environment, chatbots are expected to offer extensive emotional support mimicking humans' social behaviors.
Prior research has uncovered potential adverse consequences associated with emotional dependency on anthropomorphic social chatbots~\cite{laestadius2022too}; they may also induce negative emotions such as guilt, feelings of neglect, and experiences of gaslighting~\cite{laestadius2022too}. 
Moreover, while a limited number of studies have explored human-chatbot relationships in longitudinal study setups (see e.g. ~\cite{skjuve2023longitudinal} on a companion social chatbot), this type of research has not been explored in the group context.
Thus, future research can explore how group processes evolve over time e.g. in multi-agent interactions, focusing on preventing harmful emotional dependency and addressing potential ethical concerns.

\subsection{Limitations}
The participants we recruited from the SMARPP program exhibited significant motivation toward their recovery efforts and demonstrated a strong commitment to the program.
We gained insights into demotivators from the perspectives of facilitator participants and early SMARPP involvement of client participants, as well as their descriptions of disengagement among their dropout peers. We acknowledge that those clients who withdrew from the program prematurely may provide valuable insights into the factors that motivate or demotivate active participation in group therapy that we were unable to capture.
However, accessing these participants from the hospital side poses challenges due to their early dropouts and potentially ongoing substance use conditions, which together raise privacy issues and practical hindrances.
Future research could seek alternative routes to reach out to these individuals for further insights.

It is worth noting that all of our study's participants, both facilitator and client participants, were recruited from Japan, which limits the generalisability of our findings to other contexts. 
While we anticipate that our results will be applicable across various cultural contexts, given that SMARPP was designed based on models from Western contexts and adheres to group practices described in Western literature, future research may expand this work by further investigating cultural factors in group therapy.
Furthermore, we investigated motivators and demotivators in group therapy in the context of substance use disorder. While our results and proposed model potentially apply to group therapy in general, given the alignment with prior medical studies on the subject, we expect future studies to expand on our findings by examining other mental health treatments for further insights.

\section{Conclusion}
In this article, we employed a semi-structured interview with a specific case study on a SMARPP program in addressing substance use disorder. 
We identified the factors that influence client engagement and examined communication among stakeholders in group therapy through the lens of SDT.
Our findings shed light on the strengths and limitations of group therapy in terms of client engagement and highlight the potential of collaborative technology to improve client engagement by taking into account the communication and motivational dynamics among stakeholders in group therapy.

\begin{acks}

We thank Toshihiko Matsumoto in National Center of Neurology and Psychiatry for his support on interview participant recruitment.
This project was partly supported by JST PRESTO (Grant Number: JPMJPR23IB), JSPS Bilateral Collaboration between Japan and Finland (Grant Number: JPJSBP 120232701), and Research Council of Finland (Grant Numbers: 354161 and 349637).

\end{acks}


\bibliographystyle{ACM-Reference-Format}
\bibliography{sample-base}
\begin{table*}[h!]
    \centering
    \caption{Interview questions for facilitator participants.}
    \label{tab:faci_q}
    \begin{tabular}{ll}
    \toprule[1pt]\midrule[0.3pt]
        Category & Question \\
        \midrule
        \multirow{3}{10em}{General experience as a facilitator in SMARPP} 
        & How many years have you worked as a facilitator, and how often? \\
        & What is the overall structure of SMARPP? \\
        & What are your responsibilities as a facilitator? \\
        \midrule
        \multirow{9}{10em}{Communication Practices} 
        & How do you approach communication with your clients during check-in/workbook 
            \vspace{-0.3em} \\ & \ \ \ \ 
            reading/check-out sessions? For instance, what questions or comments do you use? \\
        & Why do you use these particular questions or comments? \\
        & What other aspects are important during check-in/workbook reading/check-out sessions,
            \vspace{-0.3em} \\ & \ \ \ \ 
            and how do you address them in communication with the client? \\
        & Can you provide an example of a client who successfully improved their engagement 
            \vspace{-0.3em} \\ & \ \ \ \ 
            and commitment to treatment through communication with you? \\
        & Can you provide an example of a client who faced challenges in communicating 
            \vspace{-0.3em} \\ & \ \ \ \ 
            with you and subsequently discontinued treatment? \\
        \midrule
        \multirow{2}{10em}{Pros and Cons of SMARPP Features} & What are the advantages and disadvantages of the in-person style of treatment? \\
        & What are the advantages and disadvantages of the group style of treatment? \\
        \midrule
        \multirow{3}{10em}{Desires in Facilitation} & What do you find challenging in your role as a facilitator? \\
        & If you had the opportunity, what changes or improvements would you make to your 
            \vspace{-0.3em} \\ & \ \ \ \ 
            facilitation approach? \\
        \midrule
        \multirow{2}{10em}{Special Considerations} & What considerations do you take into account when working with first-time SMARPP 
            \vspace{-0.3em} \\ & \ \ \ \ 
            participants? \\
        \bottomrule[1pt]
    \end{tabular}
\end{table*}

\begin{table*}
    \centering
    \caption{Interview questions for client participants.}
    \label{tab:pat_q}
\begin{tabular}{ll}
        \toprule[1pt]\midrule[0.3pt]
        Category & Question \\
        \midrule
        \multirow{2}{10em}{General Experience as a SMARPP Client} & How many years have you participated in the program as a client, and how often? \\
        & What is the overall structure of SMARPP from a client's perspective? \\
        \midrule
        \multirow{2}{10em}{Effects of Communication} & Have you had positive or negative experiences when communicating with a facilitator? \\
        & Have you had positive or negative experiences when communicating with other participants? \\
        \midrule
        \multirow{2}{10em}{Pros and Cons of SMARPP Features} & What are the advantages and disadvantages of the in-person style of treatment? \\
        & What are the advantages and disadvantages of the group style of treatment? \\
        \midrule
        \multirow{3}{10em}{Changes in Mental State} & How did you feel when you first participated in the program? \\
        & How have you noticed changes in your mental state during your participation in the program? \\
        & What factors or aspects do you believe have contributed to your treatment progress? \\
        \bottomrule[1pt]
    \end{tabular}
\end{table*}

\newpage
\appendix
\section{Guide Questions for Interviews}
Table~\ref{tab:faci_q} and Table~\ref{tab:pat_q} show examples of questions asked in our interview for facilitator and client participants correspondingly. The actual interviews followed the semi-structured interview format, and questions were added or modified as needed according to the participants' answers.
\clearpage

%

\end{document}